\documentclass[aps, pre, showpacs,twocolumn,floatfix,10pt]{revtex4-1}

\usepackage{graphicx} 
\usepackage{amsmath,amssymb,amsthm}

\usepackage{xfrac}			
\usepackage{microtype}		
\usepackage{color}			
\newtheorem*{lem}{Lemma}	

\let\oldmarginpar\marginpar
\renewcommand\marginpar[1]{\-\oldmarginpar[\raggedleft\footnotesize #1]%
{\raggedright\footnotesize #1}}

\begin{document}

\title{Virtual potentials for feedback traps}

\author{Yonggun Jun and John Bechhoefer}

\affiliation{Department of Physics, Simon Fraser University, Burnaby, B.C., V5A 1S6, Canada}

\date{\today} 

\begin{abstract}
The recently developed feedback trap can be used to create arbitrary virtual potentials, to explore the dynamics of small particles or large molecules in complex situations.  Experimentally, feedback traps introduce several finite time scales:  there is a delay between the measurement of a particle's position and the feedback response;  the feedback response is applied for a finite update time; and a finite camera exposure integrates motion.  We show how to incorporate such timing effects into the description of particle motion.  For the test case of a virtual quadratic potential, we give the first accurate description of particle dynamics, calculating the power spectrum and variance of fluctuations as a function of feedback gain, testing against simulations.  We show that for small feedback gains, the motion approximates that of a particle in an ordinary harmonic potential.  Moreover, if the potential is varied in time, for example by varying its stiffness, the work that is calculated approximates that done in an ordinary changing potential.  The quality of the approximation is set by the ratio of the update time of the feedback loop to the relaxation time of motion in the virtual potential.
\end{abstract}


\maketitle

\section{Introduction}

In 2005, Cohen and Moerner introduced the Anti-Brownian ELectrokinetic (ABEL) trap, a new experimental technique for studying long-time dynamical properties of small particles and molecules.  One key advantage over other trapping techniques  such as optical or magnetic tweezers is its ability to trap molecules and sub-micron particles directly rather than via the micron-sized particles of the former techniques.  The ABEL trap uses feedback to counteract the random thermal fluctuations that perturb the motion of small objects in a finite-temperature fluid \cite{cohen05a}, with electrokinetic forces being one way among many to apply restoring forces; it is thus perhaps more simply termed a \textit{feedback trap}.  The basic idea is to observe the position of an object, compare its estimated position with a desired position, and, as rapidly as possible, apply a corrective force to move the particle towards the desired position.  To trap small objects, the observations must be rapid, as the particle diffuses ``out of control" during the time $t_s$ between corrections.  Indeed, the lower limit to the size of particle that can be trapped depends directly on $t_s$.  Recently, Fields and Cohen, by responding to every detected photon, were able to trap for several seconds a single fluorescent dye molecule diffusing in water \cite{fields11}.

Cohen has also shown that a feedback trap can do more than just trap a particle:  it is also possible to place the particle in an arbitrary ``virtual" potential \cite{cohen05b}.  The protocol to approximate motion in a potential $U(x)$ works as follows:  Let the estimated position of the particle at time step $n$, in units of a sampling period $t_s$, be $\bar{x}_n$.  (We distinguish between the observed position $\bar{x}_n$ and the true position $x_n$.)   Then, at time step $n$, we apply the force $F_n = -\partial_x U(\bar{x}_n)$, held constant over the interval $t_s$.  In \cite{cohen05b}, experimental evidence is given that the probability distribution of observed positions $\bar{x}$ obeys the Boltzmann distribution $\rho(\bar{x}) \propto \exp[-U(\bar{x})/k_BT]$.  

In the above scheme, the potential imposed is a virtual one that is imposed by the rules of the feedback loop.  Take away the feedback loop, and there is only a particle diffusing in a fluid.  But then it is fair to ask, In what sense is the motion of a discrete closed-loop feedback system equivalent to a ``real" potential?  After all, the dynamics is discretized at a time scale $t_s$.  Reasoning by analogy to computer simulations of the Langevin equation, we expect that as $t_s \to 0$, the closed-loop feedback system will be equivalent to the desired continuous dynamical system.  But experiments are always done at finite $t_s$, and the information is acted upon only after a delay, $t_d$.  In addition, the position is typically measured (with random error) by integrating a camera over a finite time $t_c$ that can be comparable to $t_s$.  Under such conditions, is the motion truly equivalent to the desired potential?  Are there corrections to the ``naive potential"?   If so, are such corrections important in a given application?  Likewise, can one use virtual potentials for thermodynamic calculations such as the work done by a changing potential?  

We will find that the answers to these questions are mathematically boring---the situation resembles that of a continuous system with discrete observations---but physically exciting:  we can now study the motion of a particle in an arbitrary potential or force field and learn about both dynamic and thermodynamic quantities.  

Indeed, this article was motivated by attempts to use a feedback trap to explore Landauer's Principle that erasing information in a memory element requires a finite amount of work  \cite{landauer61,berut12}.  After making preliminary measurements reporting a qualitative observation of the effect \cite{cho11}, we realized that there were systematic deviations from calculations based on a continuous potential that needed to be understood before a quantitative study could be made.  The calculations reported below address these concerns.

\section{Dynamics of a particle in a quadratic virtual potential}
\label{sec:equiv2cont}

In this section, we explore the dynamics of a feedback trap with an imposed quadratic virtual potential.  The goal will be to calculate the power spectrum and variance of the particle's position fluctuations.  The calculation is complicated by the presence of three short time scales, which are comparable but not in general equal:
\begin{enumerate}
\item The output is updated after an observation with a delay time $t_d$ . 
\item The observation is made via a camera exposure of duration $t_c$.
\item The update is applied for a time $t_s$, which is also the periodicity of camera exposures (one exposure starts every $t_s$, which may be longer than $t_c$).
\end{enumerate}
As we will see, previous work has not described all the consequences of these experimental complications.  Below, we will see how to incorporate them by solving a series of increasingly complicated problems:  
\begin{enumerate}
\item The diffusion dynamics of a Brownian particle with observations every $t_s$ that are integrated over $t_c$, with a response that is delayed by a time $t_d=t_s$; 
\item feedback trapping in a virtual quadratic potential, with noiseless, instantaneous position measurements, whose results are available with no delay;
\item feedback trapping adding a camera exposure $t_c$ and a delay $t_d=t_s$; 
\item feedback trapping in the general case where $t_d \neq t_s$.
\end{enumerate}

\subsection{Free diffusion}
\label{sec:free-diffusion}

\begin{figure}[ht]
	\centering\includegraphics[width=3.0in]{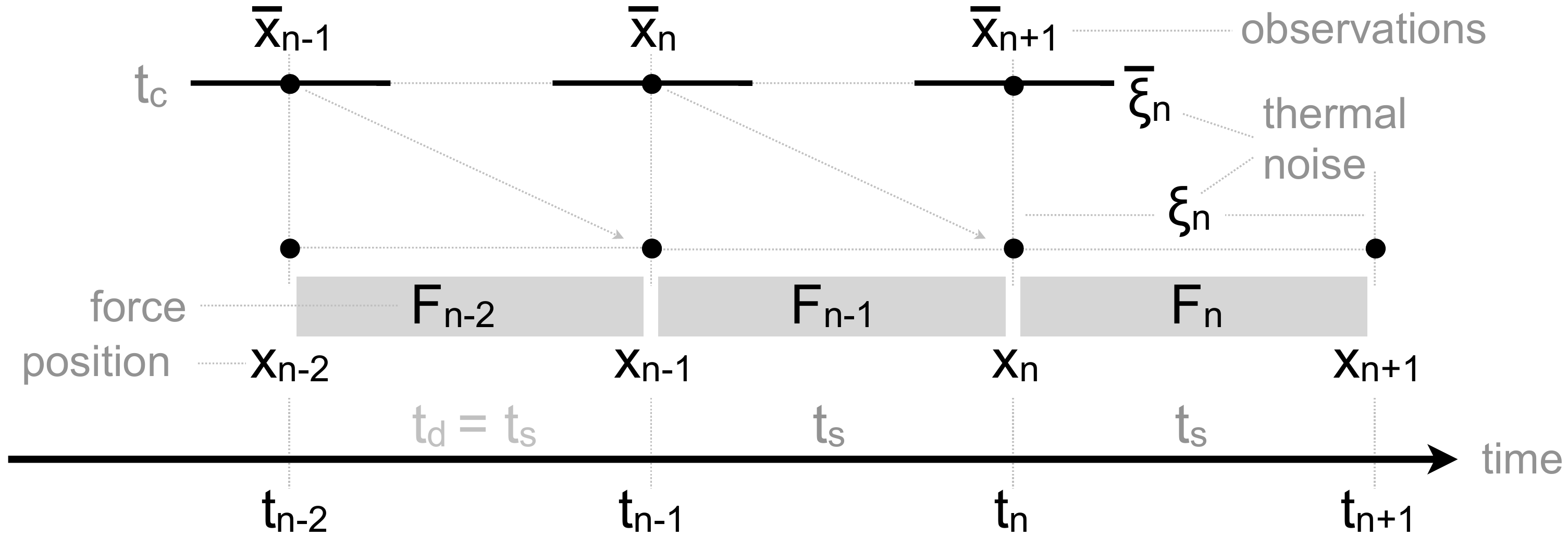}
	\caption{Timing diagram for feedback trap with instantaneous position measurements.  The feedback update interval is $t_s$, and there is a delay $t_d=t_s$ between the measurement time and the output of the next feedback update.  The true position of particles $x_n$ are indicated at bottom, together with the associated thermal noise $\xi_n$ and force $F_n$.  The force $F_n$ is computed from the observed position $\bar{x}_n$ available at the time $F_n$ is started, $nt_s$.  At top are indicated the observed position $\bar{x}_n$ and the amount of associated thermal noise $\bar{\xi}_n$ in each observation.}
\label{fig:timing-diffusion}
\end{figure}

We begin by considering the one-dimensional free diffusion of a Brownian particle that is observed via camera exposures that measure the average position over the exposure time $t_c$.  There is one exposure every $t_s$ interval, and the mid-point of the exposure time is delayed by one time step, $t_d=t_s$.  We distinguish between the actual position of a particle at time $n t_s$, denoted by $x_n$ and the corresponding observation of that position (as deduced from the camera exposure), denoted $\bar{x}_n$.  The above statements are illustrated in the timing diagram shown in Fig.~\ref{fig:timing-diffusion}.

We start with the equation of motion of the Brownian particle, assuming complete overdamping and no applied forces: 
\begin{equation}
	\dot{x} = \frac{1}{\gamma} \, \xi^{(F)}(t) \equiv \xi^{(v)}(t) \,,
\label{eq:continuous-eq-of-motion}
\end{equation}
where $\xi^{(F)}(t)$ is the fluctuating thermal force and $\xi^{(v)}(t)$ gives the corresponding velocity fluctuations.  From the fluctuation-dissipation theorem \cite{kubo66, marconi08}, $\langle \xi^{(v)}(t) \rangle = 0$ and $\langle \xi^{(v)}(t) \, \xi^{(v)}(t') \rangle = 2D \, \delta(t-t')$, where the diffusion constant $D = k_BT/\gamma$ and $\delta(t-t')$ is the Dirac delta function.  Integrating the continuous equation of motion over the time interval $[n t_s, (n+1) t_s)$ then gives
\begin{equation}
	x_{n+1} = x_n + \xi_n \,, \qquad \xi_n = 
	\int_{nt_s}^{(n+1)t_s} \xi^{(v)}(t) \, dt \,.
\label{eq:discrete-dynamics-general1}
\end{equation}
where $\xi_n$ represents the displacement due to thermal forces integrated over time interval $n$.  Again, $\xi_n$ are Gaussian random variables with mean 0 and with $\langle \xi_m \, \xi_n \rangle = 2Dt_s \delta_{mn}$, where  $\delta_{mn}$ is the Kronecker delta function. 

We next consider the distinction between the actual position $x_n$ and the observed position $\bar{x}_n$ that becomes available at the same time.  Taking into account that the observed position is based on an exposure of duration $t_c$ whose midpoint is delayed by $t_d=t_s$, we have
\begin{align}
	\bar{x}_{n+1} &= \frac{1}{t_c} 
	\int_{nt_s-\frac{1}{2}t_c}^{nt_s+\frac{1}{2}t_c} x(t) \, dt
	= x_n - \xi^{(0)}_n + \bar{\xi}_n \,, \nonumber \\
	\xi^{(0)}_n &= \int_{nt_s-\frac{1}{2}t_c}^{n t_s} \xi^{(v)}(t) \, dt \,,
		\nonumber \\
	\bar{\xi}_n &= \frac{1}{t_c} \int_{nt_s-\frac{1}{2}t_c}^{nt_s+\frac{1}{2}t_c}
		dt \, \int_{nt_s-\frac{1}{2}t_c}^t \xi^{(v)}(t') \, dt' \,.
\label{eq:obs-pos}
\end{align}
In Eq.~\eqref{eq:obs-pos},  $x_n - \xi_n(0)$ is the position of the particle at the beginning of the exposure.  The term $\bar{\xi}_n$ represents the thermal noise as averaged by the camera exposure and gives the average displacement from the position at the beginning of the exposure to its end.  The negative $\xi^{(0)}_n$ is needed because we define the measured position $\bar{x}_{n+1}$ relative to the position at the midpoint of the camera exposure and not at the beginning.  (This convention will be convenient for what follows.)

Next, we consider the statistics of the measured particle displacement 
\begin{align}
	\overline{\Delta x}_n &\equiv \bar{x}_{n+1}-\bar{x}_n \nonumber \\
	&= \xi_{n-1} - \xi^{(0)}_n + \xi^{(0)}_{n-1} + \bar{\xi}_n - 
		\bar{\xi}_{n-1} \nonumber \\
	&= \xi'_{n-1} + \bar{\xi}_n - \bar{\xi}_{n-1} \,.
\label{eq:displacement1}
\end{align}
Because the thermal noise terms in Eq.~\eqref{eq:obs-pos} are all Gaussian with mean zero, we immediately have that $\langle \overline{\Delta x}_n \rangle = 0$, which simply says that there is no bias to the displacements of a random walker.  In the last line of Eq.~\eqref{eq:displacement1},
\begin{align}
	\xi'_{n-1}  \equiv \xi_{n-1} - \xi^{(0)}_n + \xi^{(0)}_{n-1} = 
	\int_{(n-1)t_s-\frac{1}{2}t_c}^{nt_s-\frac{1}{2}t_c} \xi^{(v)}(t) \, dt \,,
\label{eq:newnoise}
\end{align}
which is just the thermal noise over one sample period, $t_s$, starting and ending at  the beginning of the camera exposure.  By contrast, $\xi_{n-1}$ is the thermal noise over one period starting and ending at the midpoint of the camera exposure.

The mean-square displacement during $t_s$ is then
\begin{align}
	\left\langle (\overline{\Delta x}_n)^2 \right\rangle 
	= \left\langle (\xi'_{n-1})^2 \right\rangle 
	+ 2 \left\langle (\bar{\xi}_n)^2 \right\rangle 
	- 2 \left\langle \xi'_{n-1} \bar{\xi}_{n-1} \right\rangle \,.
\label{eq:msd}
\end{align}
In Eq.~\eqref{eq:msd}, we omit the terms $2 \langle \xi'_{n-1} \bar{\xi}_n\rangle = -2\langle \bar{\xi}_{n-1} \, \bar{\xi}_n \rangle = 0$ (no overlap).  We can evaluate
\begin{align}
	 \langle \xi'_{n-1} \bar{\xi}_{n-1} \rangle &= \frac{1}{t_c}
	 \left \langle \int_0^{t_s} \xi^{(v)}(t'') \, dt'' \int_0^{t_c} dt 
	 	\int_0^{t}  \xi^{(v)}(t') \, dt' \right \rangle \nonumber \\
	&= \frac{1}{t_c}  \int_0^{t_c} dt  \int_0^{t} dt' \int_0^{t_s}
	 \left\langle \xi^{(v)}(t') \, \xi^{(v)}(t'') \right\rangle \, dt'' \nonumber \\[3pt]
	&= \frac{2D}{t_c} \int_0^{t_c} dt  \int_0^{t} dt' 
		\int_0^{t_s} \delta(t'-t'') \, dt'' \nonumber \\[3pt]
	&=  \frac{2D}{t_c} \int_0^{t_c} dt  \int_0^{t} (1) \, dt' \nonumber \\
	&= \frac{2D}{t_c} \int_0^{t_c} t \, dt \nonumber \\
	&= Dt_c \,.
\label{eq:cross-corr}
\end{align}
where we have shifted the domain of integration for all the integrals by $(n-1)t_s - \tfrac{1}{2} t_c$, for clarity.  A similar calculation by Cohen \cite{cohen06} gives $\langle (\bar{\xi}_n)^2 \rangle = \tfrac{2}{3} Dt_c $.  Putting these results together, we have
\begin{align}
	\langle (\overline{\Delta x}_n)^2 \rangle &= 2Dt_s + 2 \left( \tfrac{2}{3} Dt_c \right) -2 (Dt_c) \nonumber \\
	&= 2D(t_s-\tfrac{1}{3}t_c) \,,
\label{eq:msd2}
\end{align}
which gives the finite-exposure-time correction to the usual mean-square displacement \cite{goulian00,savin05,cohen06}.

Turning now to the power spectrum of the position measurements, we take the $Z$-transform (or equivalently, calculate the generating function) of Eqs.~\eqref{eq:discrete-dynamics-general1} and \eqref{eq:obs-pos}.  Let us define the $Z$-transform of the sequence $x_n$ to be
\begin{equation}
	\mathcal{Z}[x_n] = x(z) \equiv \sum_{n=0}^\infty x_n z^{-n} \,,
\end{equation}
with similar definitions for all other quantities ($\bar{x} = \mathcal{Z}(\bar{x}_n)$, etc.).  Neglecting initial conditions, we then have
\begin{subequations}
\begin{align}
	\label{eq:ztrans1a}
	(z-1) x &= \xi  \\	
	z \bar{x} &= x - \xi^{(0)} + \bar{\xi} \,.
	\label{eq:ztrans1b}
\end{align}
\end{subequations}
Multiplying Eq.~\eqref{eq:ztrans1b} by $z-1$ and substituting Eq.~\eqref{eq:ztrans1a} gives
\begin{align}
	(z-1) z \bar{x} &= \xi - (z-1) \xi^{(0)} + (z-1) \bar{\xi} \nonumber \\
	&\equiv \xi'+ (z-1) \bar{\xi} \,,
\end{align}
where we make the same redefinition of the thermal noise term that we did in Eq.~\eqref{eq:newnoise}, with $\xi'(z)$ the $Z$-transform of the thermal noise $\xi'_n$.  The power spectrum (modulus of discrete-time Fourier transform) is then
\begin{align}
	|z(z-1)|^2 \left\langle |\bar{x}|^2 \right\rangle 
	 &= \nonumber \\
	2 \left[ \left\langle |\xi'|^2 \right\rangle 
	+ |z-1|^2 \left\langle |\bar{\xi}|^2 \right\rangle 
	+ (z-1) \left\langle \xi' \, \bar{\xi} \right\rangle + c.c. \right] \,,
\end{align}
where $c.c.$ denotes the complex conjugate of the last terms and where the factor of 2 results from considering only positive frequencies.  Evaluating the thermal noise expressions and substituting $z=e^{i\omega t_s}$, we have
\begin{align}
	|z(z-1)|^2 \left\langle |\bar{x}|^2 \right\rangle &= 2 [ 2Dt_s 
	+ 2(1-\cos \omega t_s) (\tfrac{2}{3}Dt_c) \nonumber \\
	&+ (e^{i\omega t_s} -1) (Dt_c) + c.c. ] \nonumber \\
	&= 4Dt_s-\tfrac{4}{3} Dt_c (1-\cos \omega t_s) \,.
\end{align}
Solving for $ \langle |\bar{x}|^2 \rangle$ as a function of the angular frequency $\omega$, we have
\begin{equation}
	 \left\langle |\bar{x}|^2 \right\rangle = 
	 \frac{2Dt_s-\tfrac{2}{3} Dt_c (1-\cos \omega t_s)}
	 {(1-\cos \omega t_s)} \,.
\end{equation}

\subsection{Harmonic potential, with ``perfect" measurements}
\label{sec:harmonic-perfect}

We begin our study of virtual potentials in the simplest case, where the measurements are ``perfect":  that is, they are instantaneous ($t_c=0$), without delay ($t_d=0$), and free of observation noise.  We maintain an update time of $t_s$.  Although no experiment is so simple, the results already illustrate some of the key differences between motion in a true and in a virtual potential.  Under these assumptions, the equations of motion Eq.~\eqref{eq:discrete-dynamics-general1} become
\begin{equation}
	x_{n+1} = x_n -\alpha x_n + \xi_n \,.
\label{eq:discrete-dynamics-perfect}
\end{equation}
We first note that Eq.~\eqref{eq:discrete-dynamics-perfect} is identical to the Euler algorithm for integrating stochastic differential equations \cite{gardiner09}.  In Eq.~\eqref{eq:discrete-dynamics-perfect}, we can view the $\alpha$ term as deriving from a ``proportional feedback" law, $F_n = -k x_n$, with $\alpha = t_s k /\gamma$.  Although the $k$ in the feedback law superficially resembles the force constant $k^{(c)}$ in Hooke's law, $F(t) = -k^{(c)} x(t)$, it is not quite the same quantity.  To see this point qualitatively, we note that the force in Hooke's law continuously changes as $x(t)$ changes, whereas the force in the feedback system $F_n$ is constant over the interval $t_s$.  As long as $\alpha \ll 1$, the difference between the two situations is not great.  But larger values of $\alpha$ lead to different dynamics.  

To illustrate this point, we calculate the steady-state variance.  Assuming $\langle x_{n+1}^2 \rangle = \langle x_n^2 \rangle \equiv \langle x^2 \rangle$,   squaring Eq.~\eqref{eq:discrete-dynamics-perfect}, and noting that $\langle x_n \xi_n \rangle = 0$, we have
\begin{align}
	 \langle x^2 \rangle = (1-\alpha)^2  \langle x^2 \rangle +  \langle \xi^2 \rangle \,,
\end{align}
where $\langle \xi_n^2 \rangle \equiv \langle \xi^2 \rangle = 2Dt_s$.  With $1-(1-\alpha)^2 = \alpha(2-\alpha)$, we then have
\begin{align}
	 \langle x^2 \rangle = \frac{2Dt_s}{\alpha (2-\alpha)}  \,.
\label{eq:x2var0delay}
\end{align}
For $\alpha \ll 1$, the variance is $Dt_s/\alpha$, which is the value expected by a ``naive" application of the equipartition principle: substituting $\alpha = t_s k /\gamma$ and $D = k_B T/\gamma$ gives $\langle x^2 \rangle = k_BT/k$.  However, the variance at finite $\alpha$ is always larger and diverges at $\alpha^*=2$, beyond which the motion is unstable.  Physically, the extra variance comes from over-correcting perturbations.  At $\alpha = 2$, the motion oscillates from one side of equilibrium to the other.

\begin{figure}[ht]
	\centering\includegraphics[width=2.3in]{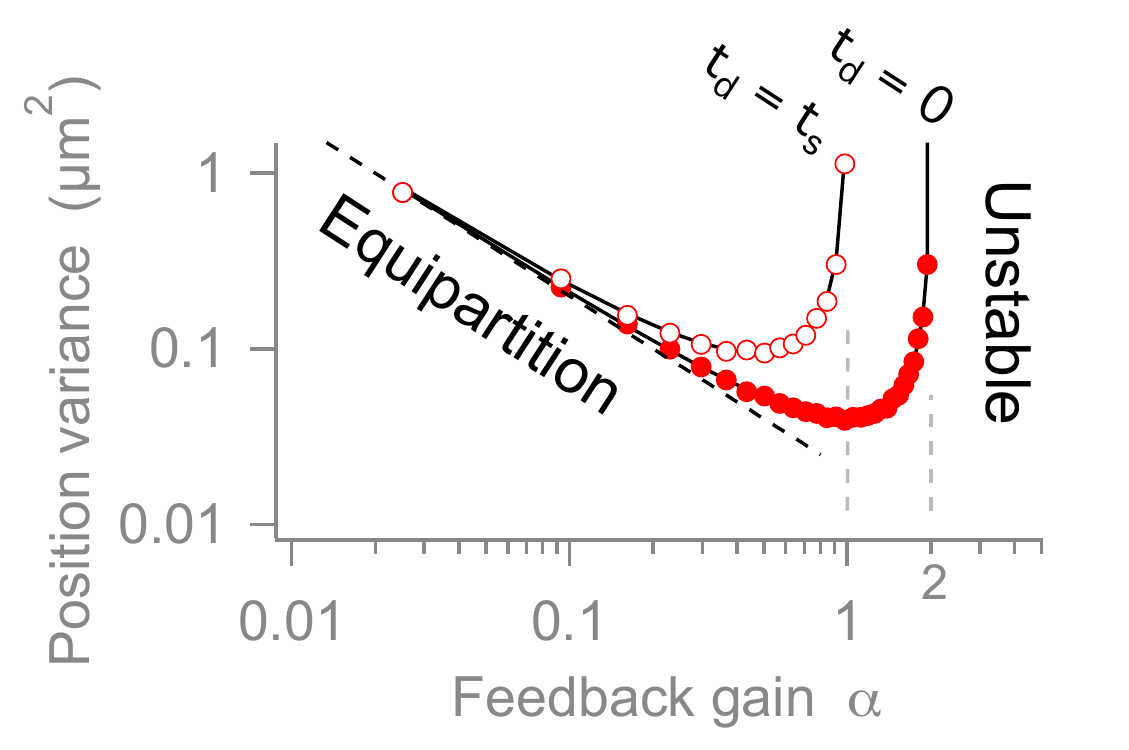}
	\caption{(Color online.)  Position variance for a virtual harmonic potential in a feedback trap, for $t_s = 0.01$ s and $D = 2~\mu$m$^2/$s.  Simulations based on Eqs.~\eqref{eq:discrete-dynamics-perfect} and \eqref{eq:discrete-dynamics-perfect1} are for the same parameters, for a total simulation time of $\tau=1000$ s.  The solid round markers correspond to the case of $t_d=0$, and the accompanying theory curve is given by Eq.~\eqref{eq:x2var0delay}.  The hollow round markers correspond to the case $t_d=t_s$, and the accompanying theory curve is given by Eq.~\eqref{eq:x2var1delay}.  The dashed line shows the ``naive" variance calculated from the equipartition theorem $\langle x^2 \rangle = (k_BT/k) = Dt_s/\alpha$.  The vertical dashed gray line indicates unstable motion, which occurs beyond $\alpha^* = 2$ for $t_d=0$ and $\alpha^*=1$ for $t_d=t_s$.}
\label{fig:x2_0}
\end{figure}

Repeating the argument for unit delay, $t_d=t_s$, we have
\begin{equation}
	x_{n+1} = x_n -\alpha x_{n-1} + \xi_n \,.
\label{eq:discrete-dynamics-perfect1}
\end{equation}
For this case, squaring and averaging Eq.~\eqref{eq:discrete-dynamics-perfect1} leads to $-\alpha^2 \langle x^2 \rangle = 2Dt_s - 2\alpha \langle x \, x_{-1} \rangle$.  The last term can be evaluated by multiplying Eq.~\eqref{eq:discrete-dynamics-perfect1} by $x_n$ and averaging, giving $\langle x \, x_{-1} \rangle = \langle x^2 \rangle - \alpha \langle  x \, x_{-1} \rangle$.  The result is
\begin{align}
	 \langle x^2 \rangle = 
	 2Dt_s \left[ \frac{1+\alpha}{\alpha(1-\alpha)(2+\alpha)} \right] \,,
\label{eq:x2var1delay}
\end{align}
which is unstable for $\alpha > 1$ and again goes to the equipartition result $Dt_s/\alpha$ when $\alpha \to 0$.  The two expressions for $\langle x^2 \rangle$, along with corresponding simulations, are plotted as a function of feedback gain $\alpha$ in Fig.~\ref{fig:x2_0}.  We note that the reduction in critical gain $\alpha^*$ from 2 to 1 reflects the performance deterioration caused by the delay.  Longer delays further decrease $\alpha^*$. For example, a similar calculation gives $\alpha^*= \tfrac{1}{2}(\sqrt{5}-1) \approx 0.62$ for $t_d=2t_s$.

\subsection{Harmonic potential, with camera exposure}
\label{sec:harmonic-uniform-delay}

We next include a finite camera exposure $t_c$ and also observation noise $\chi_n$.  The equations of motion generalize in two ways.  First, we define the force $F_n$ to act between time $nt_s$ and $(n+1)t_s$.  As Fig.~\ref{fig:timing-diffusion} shows, the force changes midway through the camera exposure time.  Integrating the equation of motion for the continuous dynamics forwards and backwards in time from $t_n$ and neglecting stochastic forces, we find, 
\begin{equation}
	x(t) = \begin{cases}
			x_n - \frac{1}{\gamma} F_{n-1} (t_n-t) & t_{n-1}<t<t_n \\[6pt]
			x_n + \frac{1}{\gamma} F_n (t-t_n) & t_n<t<t_{n+1} \,.			\end{cases}
\label{eq:xt-cases}
\end{equation}
Inserting the expression for $x(t)$ in Eq.~\eqref{eq:xt-cases} into the average for $\bar{x}_{n+1}$ in Eq.~\eqref{eq:obs-pos}, we have
\begin{align}
	\bar{x}_{n+1} &= \frac{1}{t_c} 
	\int_{t_n-\frac{1}{2}t_c}^{t_n} 
	\left[ x_n - \frac{1}{\gamma} F_{n-1} (t_n-t) \right] \, dt \nonumber \\
	 & \qquad + \frac{1}{t_c}  \int_{t_n}^{t_n+\frac{1}{2} t_c}  \left[
	 x_n + \frac{1}{\gamma} F_n (t-t_n) \right]  \, dt  \nonumber \\
	 &= x_n + \left( \frac{F_{n-1}}{\gamma t_c} \right) \int_{-\frac{1}{2}t_c}^0 t \, dt
	 + \left( \frac{F_n}{\gamma t_c} \right) \int_0^{\frac{1}{2}t_c} t \, dt \nonumber \\
	 &= x_n + \frac{t_c}{8\gamma} (F_n-F_{n-1}) \,.
\label{eq:xbias}
\end{align}
Equation~\eqref{eq:xbias} implies that the measured position is biased by a difference in forces in intervals $n$ and $n+1$.  In particular, there will be no bias if the forces are constant, $F_{n-1}=F_n$.  In that case, there is a uniform drift; the midpoint of the exposure is at $x_n$; and the particle moves, on average, an equal amount before and after the midpoint (in time) of the exposure.

Taking into account the unit delay $t_d=t_s$ leads to coupled, linear difference equations for $x_n$ and $\bar{x}_n$:
\begin{align}
	x_{n+1} &= x_n - \alpha \bar{x}_n 
		+ \xi_n \nonumber \\
	\bar{x}_{n+1} &= x_n + \alpha' (\bar{x}_{n-1}-\bar{x}_n) 
		- \xi^{(0)}_n + \bar{\xi}_n + \chi_n \,,
\label{eq:discrete-dynamics-harmonic}
\end{align}
where $\alpha = \frac{t_s}{\gamma/k}$ is the ratio of the sampling time to the trap relaxation time, $\alpha' = \alpha \tfrac{t_c}{8t_s}$ measures the effects of the finite camera exposure time $t_c$, and where we have included stochastic terms in the $\bar{x}$ equation.  In particular, $\chi_n$ accounts for observation noise due to photon shot noise, microscope resolution, and other effects.  For simplicity, we assume it to be Gaussian, with $\langle \chi_n \rangle = 0$ and $\langle \chi_n^2 \rangle = \chi^2$.  The distinction between the state variable (the position $x_n$) and its measurement $\bar{x}_n$ in Eq.~\eqref{eq:discrete-dynamics-harmonic} is emphasized in discussions of control theory \cite{astrom97,bechhoefer05}.

To find the power spectrum, we generalize slightly the analysis of Sec.~\ref{sec:free-diffusion}.  Taking the $Z$-transform, we have
\begin{align}
	(z-1) x = -\alpha \bar{x} + \xi \,,
\end{align}
which implies
\begin{widetext}
\begin{align}
	(z-1)z \bar{x} &= -\alpha \bar{x} -\alpha' \left(\frac{(z-1)^2}{z} \right) 
	\bar{x} + \xi' + (z-1) (\bar{\xi} + \chi) \nonumber \\[3pt]
	[z^2-(1-\alpha')z + (\alpha-2\alpha') + \alpha' z^{-1}] \bar{x} &=
	\xi' + (z-1) (\bar{\xi} + \chi) \,.
\end{align}
Solving for $\bar{x}$ then leads to an expression for the power spectrum:
\begin{subequations}
\label{eq:powerspec1}
\begin{align}
	\bar{x} &= \frac{\xi' + (z-1) (\bar{\xi} + \chi)}
	{[z^2-(1-\alpha')z + (\alpha-2\alpha') + \alpha' z^{-1}]}  
\label{eq:powerspec1a} \\[6pt]
	  \left\langle |\bar{x}|^2 \right\rangle &= 
	 \frac{2 \left[ 2Dt_s + 2 (\chi^2 - \tfrac{1}{3} Dt_c) 
	 (1-\cos \omega t_s) \right]}
	 {|e^{2i\omega t_s} - (1-\alpha') e^{i\omega t_s} + (\alpha - 2\alpha') 
	 + \alpha' e^{-i\omega t_s} |^2 } \,.
\label{eq:powerspec1b}
\end{align}
\end{subequations}
\end{widetext}

To our knowledge, Eq.~\eqref{eq:powerspec1} has not been previously derived.  Previous versions \cite{gosse02, cohen06} neglect the $\alpha'$ terms in the denominator.  Physically, those terms are present because the averaged position $\bar{x}_n$ and not $x_n$ is used in the feedback loop.  In general, the denominator in transfer functions such as Eq.~\eqref{eq:powerspec1a}  reflects the structure of feedback loops \cite{bechhoefer05,astrom97}.

\begin{figure}[ht]
	\centering\includegraphics[width=2.3in]{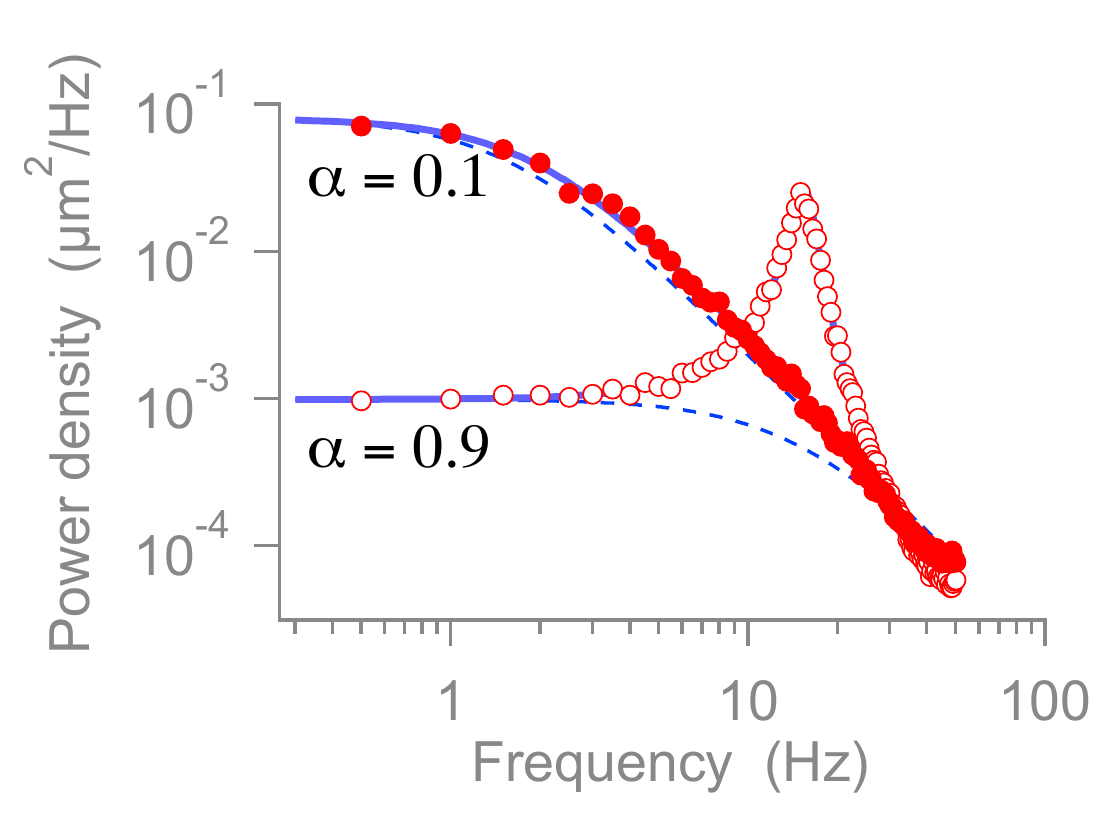}
	\caption{(Color online.)  Power spectra for a virtual harmonic potential in a feedback trap, for $t_s = 0.01$ s, $(t_d/t_s) = 1$, $(t_c/t_s) = 0.95$, $D = 2~\mu$m$^2/$s, $\chi = 0.018~\mu$m, and $\alpha = 0.1$ and 0.9.  Simulations are for the same parameters, for simulation time $\tau=400$ s.  The power spectra (solid lines) are plotted from Eq.~\eqref{eq:powerspec1b} and are not fit to the simulations.  The corresponding Lorentzian approximations, Eq.~\eqref{eq:lor-approx}, are shown as dashed lines.}
\label{fig:PSgraphs}
\end{figure}

In Fig.~\ref{fig:PSgraphs}, we illustrate typical power spectra for low ($\alpha = 0.1$) and high ($\alpha = 0.9$) values of the feedback gain.  Markers denote simulations using parameter values for $D$, $t_c$, $t_d$, $t_s$, and $\chi$ that are typical of experimental systems.  The $\alpha=0.1$ case approximates the Lorentzian spectrum expected of a continuous system, although even in that case, the discrete observations ensure significant deviations from the continuous spectrum for frequencies comparable to the Nyquist frequency (50 Hz here).  The dashed line shows the corresponding Lorentzian spectrum, which, in our notation, is given by 
\begin{align}
	\langle |x|^2 \rangle = \frac{4Dt_s^2}{\alpha^2 + (\omega t_s)^2}
	= \frac{4D}{(1/t_r^2)+\omega^2} \,.  
\label{eq:lor-approx}
\end{align}	
On the other hand, the high-gain curve shows a significant resonant peak.  Physically, the peak corresponds to an overshoot.  Unlike a continuous quadratic potential, the force applied in a feedback trap is constant over the interval $t_s$.  For large feedback gains, the particle can overshoot the set point, leading to decaying damped oscillations and a resonance in the power spectrum.   The corresponding Lorentzian spectrum (dashed line) agrees with our calculation at low frequencies but then excludes the power associated with the resonance.  The instability threshold ($\alpha^* \approx 1.14$) is slightly larger than for $t_c=0$, where $\alpha^* = 1$.  The camera exposure acts as a low-pass filter of the position measurement, stabilizing the feedback loop.

\begin{figure}[ht]
	\centering\includegraphics[width=2.3in]{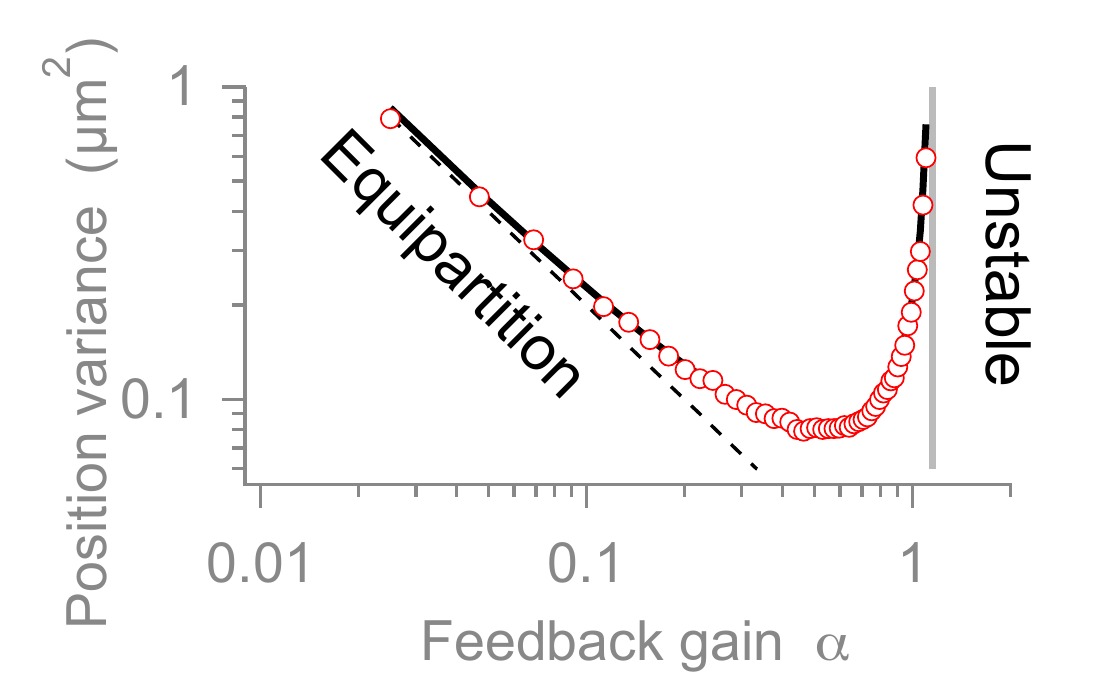}
	\caption{(Color online.)  Position variance for a virtual harmonic potential in a feedback trap, for $t_s = 0.01$ s, $(t_d/t_s) = 1$, $(t_c/t_s) = 0.95$, $D = 2~\mu$m$^2/$s, $\chi = 0.018~\mu$m.  Simulations (hollow round markers) are for the same parameters, for simulation time $\tau=400$~s.  The predicted values (solid black curve) are evaluated by integrating Eq.~\eqref{eq:powerspec1b} over frequency and are not fit to the simulations.  The dashed line shows the ``naive" variance calculated from the equipartition theorem $\langle x^2 \rangle = (k_BT/k) = Dt_s/\alpha$.  The vertical gray line indicates unstable motion for $\alpha > \alpha^* \approx 1.14$.}
\label{fig:x2}
\end{figure}

In Fig.~\ref{fig:x2}, we illustrate the simulated and predicted variance of position measurements as a function of the feedback gain $\alpha$.  Although the variance could be calculated using strategies similar to those used above for the $t_c=0$ case, it is simpler to integrate the power spectrum in Eq.~\eqref{eq:powerspec1b} from 0 to the Nyquist frequency ($2/t_s$).  For $\alpha \ll 1$, the ``naive variance" predicted from equipartition is a good approximation to the exact value.  At $\alpha = 0.1$, the delay increases the variance by about 15\%.  

\subsection{Harmonic potential, general delay}
\label{sec:harmonic-general-delay}

With zero delay, the discrete equation of motion for a particle in a feedback trap with harmonic potential is a first-order difference equation, Eq.~\eqref{eq:discrete-dynamics-perfect}.  A delay of $(t_d/t_s) = 1$, in turn, leads to a second-order difference equation, Eq.~\eqref{eq:discrete-dynamics-perfect1}.  Since an $n$th-order difference equation is the discrete analog of an $n$th-order differential equation, does a fractional delay, $(t_d/t_s) \neq$ integer, lead to an analog of a fractional derivative?  Such derivatives have an analytic structure that is qualitatively different from that of integer derivatives \cite{west03}.  In fact, the answer is much less exotic, and the results for fractional delay are qualitatively similar to those for integer delay.

\begin{figure}[ht]
	\centering\includegraphics[width=3.0in]{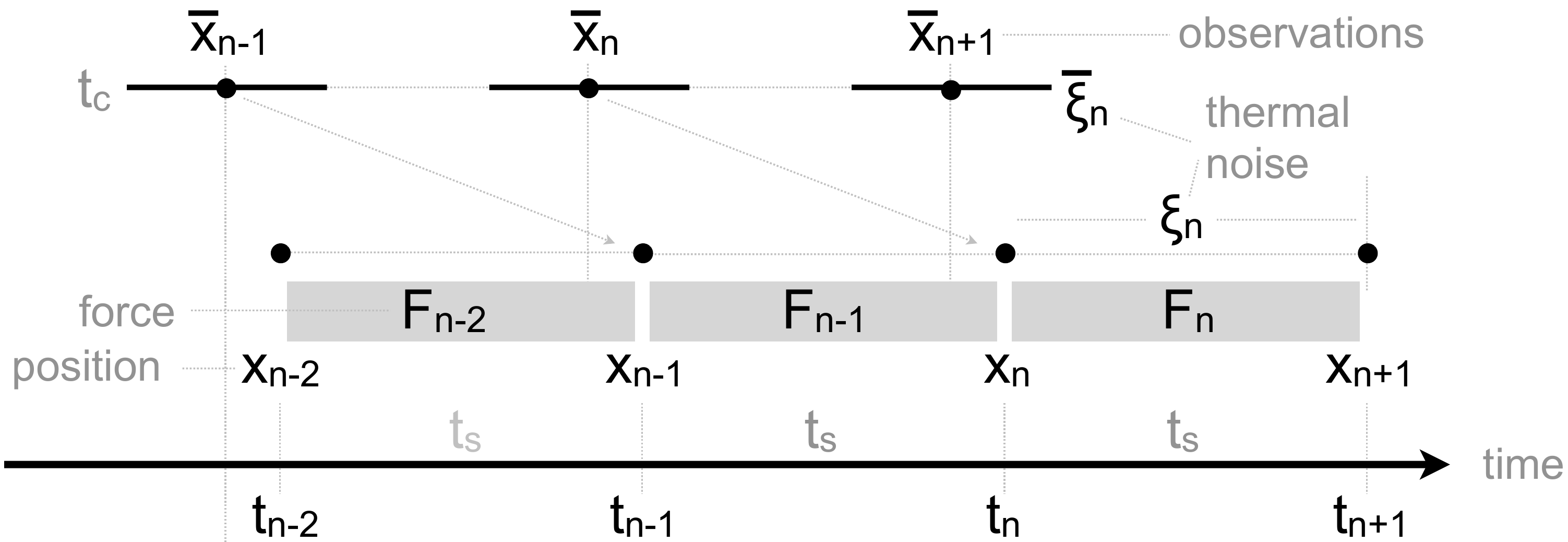}
	\caption{Timing diagram for feedback trap with instantaneous position measurements.  The timing is similar to Fig.~\ref{fig:timing-diffusion}, except that the delay time $t_d$ is not equal to the sampling time $t_s$.}
\label{fig:timing-diffusion2}
\end{figure}

To see this, we will assume that $t_d \approx t_s$.  In particular, we will assume, as illustrated in the modified timing diagram, Fig.~\ref{fig:timing-diffusion2}, that the measured position $\hat{x}_{n+1}$ is computed from a camera exposure that straddles the contributions from two forces, $F_{n-1}$ and $F_n$.  To generalize to completely arbitrary time delays requires a somewhat awkward notation that separates the number of integer periods of $t_s$ contained in $t_d$ and its remaining fractional part and also distinguishes between the straddling case treated here and the non-straddling case (similar to and simpler than the one treated here).

The first step is to repeat the derivation of the bias in $\bar{x}_{n+1}$.  The difference from Eq.~\eqref{eq:xbias} is that the integration limits now are $[t_{n+1}-t_d - \tfrac{1}{2} t_c, \, t_n)$ and $[t_n, \, t_{n+1} - t_d + \tfrac{1}{2} t_c)$.  We now find
\begin{align}
	\bar{x}_{n+1} &= x_n + 
	\frac{t_c}{8\gamma} \left( a_+ F_n  - a_- F_{n-1}  \right) \,,
\label{eq:xbias2}
\end{align}
where
\begin{align}
	a_{\pm} = \left[1 \pm \frac{2(t_s-t_d)}{t_c} \right]^2 \,.
\end{align}
Note that $a_{\pm}=1$ when $t_d=t_s$, and Eq.~\eqref{eq:xbias2} reduces to Eq.~\eqref{eq:xbias}.  The apparent divergence at $t_c \to 0$ is an artifact of our assumption that the exposure straddles two forces, which is not true in this limit.

The coupled equations for $x_n$ and $\bar{x}_n$ become
\begin{align}
	x_{n+1} &= x_n - \alpha \bar{x}_n 
		+ \xi_n \nonumber \\
	\bar{x}_{n+1} &= x_n + \alpha' (a_-\bar{x}_{n-1}-a_+\bar{x}_n) 
		- \xi^{(0)}_n + \bar{\xi}_n + \chi_n \,.
\label{eq:discrete-dynamics-harmonic2}
\end{align}
Taking the $Z$-transform then gives the transfer function and power spectrum:
\begin{widetext}

\begin{subequations}
\label{eq:powerspec2}
\begin{align}
	\bar{x} &= \frac{\xi' + (z-1) (\bar{\xi} + \chi)}
	{\{z^2-(1-\alpha' a_+)z + [\alpha-\alpha'(a_+ + a_-)] + \alpha' a_- z^{-1}\}}  
\label{eq:powerspec2a} \\[12pt]
	 \left\langle |\bar{x}|^2 \right\rangle &= 
	 \frac{2 \left[ 2Dt_s + 2 (\chi^2 - \tfrac{1}{3} Dt_c) 
	 (1-\cos \omega t_s) \right]}
	 {|e^{2i\omega t_s} - (1-\alpha' a_+) e^{i\omega t_s} 
	 + [\alpha - \alpha' (a_++a_-)]  + \alpha' a_- e^{-i\omega t_s} |^2 } \,.
\label{eq:powerspec2b}
\end{align}
\end{subequations}
\end{widetext}
As promised, the power spectrum and related results are just slightly more complicated than for integer delay.  Indeed, it is generally true that fractional delays merely change coefficients in a discrete dynamical system \cite{astrom97}.

\section{Virtual potentials and thermodynamics}
\label{sec:therm}

Can virtual potentials be used for thermodynamic calculations?  In this section, we investigate the accuracy of ``naive" calculations of the work done by a changing potential, following ideas of stochastic thermodynamics \cite{sekimoto97,sekimoto10,blickle06}.  As an example calculation, we calculate the mean work required to vary the stiffness of a virtual harmonic potential in a finite time.  We will find that estimates of work agree, to $\mathcal{O}(\alpha)$, with those of a true potential.

\subsection*{Harmonic potential with varying force constant}
\label{sec:harmonic-variable-k}

To explore the work done by a virtual potential, we consider a time-dependent potential $U(x,t)$.   We start with the case of a quadratic virtual potential with a feedback gain $\alpha_n$ that is increased in constant steps from $\alpha_i$ at $t=0$ to $\alpha_f$ at $t=\tau = Nt_s$.  Recall that $\alpha_n$ corresponds to a force $F_n = -k_n x_n$, but $k_n$ only approximates the force constant of a harmonic potential.  

To begin, we recall the calculation of the average work done in the continuous case, where the true force constant $k^{(c)}(t)$ varies from $k^{(c)}_i$ to $k^{(c)}_f$.  We further assume that the variation is done slowly enough ($\tau \to \infty$) that, at each moment, the system is in local equilibrium.  Then,
\begin{equation}
	W = \int \frac{\partial U}{\partial t} \, dt = \frac{1}{2} \int \dot{k}^{(c)} x^2 \, dt \,.
\label{eq:work-cont}
\end{equation}
Assuming that $\langle x^2 \rangle(t) = k_BT/[k^{(c)}(t)]$, we have
\begin{equation}
	\frac{\langle W \rangle}{k_BT} 
	= \frac{1}{2} \int \frac{\dot{k}^{(c)}}{k^{(c)}} dt
	=  \frac{1}{2} \int \frac{dk^{(c)}}{k^{(c)}} 
	= \frac{1}{2} \ln \frac{k^{(c)}_f}{k^{(c)}_i} \,.
\label{eq:work-cont1}
\end{equation}
For fixed time $t$, we define the partition function $\mathcal{Z}(t) = \int_{-\infty}^\infty \exp[-U(x,t)/k_BT] \, dx \sim k^{-1/2}$.  Then, in terms of the free energy $\mathcal{F} = -k_BT \ln \mathcal{Z}$, we have simply
\begin{equation}
	\langle W \rangle = \Delta \mathcal{F} \,,
\label{eq:work-cont2}
\end{equation}
as expected for an adiabatic protocol.  The calculation in Eq.~\eqref{eq:work-cont1} can be generalized to finite $\tau$ \cite{schmiedl07}.

Consider next the equivalent calculation for a feedback system with $t_d = t_c = \chi = 0$ (no delay, no camera exposure, no observation noise).   Discretizing Eq.~\eqref{eq:work-cont} gives
\begin{equation}
	\frac{\langle W \rangle}{k_BT} 
	= \frac{1}{2} \sum_n \frac{\dot{k}_n}{k_BT} \langle x_n^2 \rangle t_s 
	= \frac{\dot{\alpha}}{2D}  \sum_n \langle x_n^2 \rangle \,,
\label{eq:work0}
\end{equation}
where $\dot{\alpha} = (\alpha_f-\alpha_i)/\tau$ is constant.  Rewriting Eq.~\eqref{eq:x2var0delay} as
\begin{align}
	 \langle x^2 \rangle = Dt_s  \left( \frac{1}{\alpha} +
		\frac{1}{2-\alpha} \right)   \,,
\label{eq:x2var0delay1}
\end{align}
we have
\begin{align}
	\frac{\langle W \rangle}{k_BT} 
	&= \frac{t_s\dot{\alpha}}{2} \sum_{n=0}^N \left( \frac{1}{\alpha_n} +
		\frac{1}{2-\alpha_n} \right) \,,
\label{eq:work1}
\end{align}
where $\alpha_n = \alpha_i + \dot{\alpha} nt_s$ is the ``staircase" feedback gain.

In the Appendix, we show that
\begin{equation}
	\lim_{\tau \to \infty} t_s \dot{\alpha} \sum_{n=0}^{\tau/t_s}
	\frac{1}{b+\alpha_n} = \ln \left( \frac{b+\alpha_f}{b+\alpha_i} \right) \,.
\label{eq:lemma}
\end{equation}
Applying this result to Eq.~\eqref{eq:work1} with $b=0$ and $-2$, we have
\begin{equation}
	\frac{\langle W \rangle}{k_BT} = 
	\frac{1}{2} \ln \left[ \left( \frac{\alpha_f}{\alpha_i} \right) \left(
	\frac{2-\alpha_i}{2-\alpha_f} \right) \right] \,.
\label{eq:work2}
\end{equation}

We can quickly generalize to the case of unit delay.  Rewriting Eq.~\eqref{eq:x2var1delay} as
\begin{equation}
	\frac{\langle x^2 \rangle}{Dt_s} = \frac{1}{\alpha} 
	+ \frac{1/3}{2+\alpha} + \frac{4/3}{1-\alpha}
\end{equation} 
leads to
\begin{equation}
	\frac{\langle W \rangle}{k_BT} = 
	\frac{1}{2} \ln \left[ \left( \frac{\alpha_f}{\alpha_i} \right) \left(
	\frac{2+\alpha_f}{2+\alpha_i} \right)^{1/3}
	\left( \frac{1-\alpha_i}{1-\alpha_f} \right)^{4/3} \right] \,.
\label{eq:work3}
\end{equation}
Equations~\eqref{eq:work2} and \eqref{eq:work3} agree with simulations.

\begin{figure}[ht]
	\centering\includegraphics[width=2.3in]{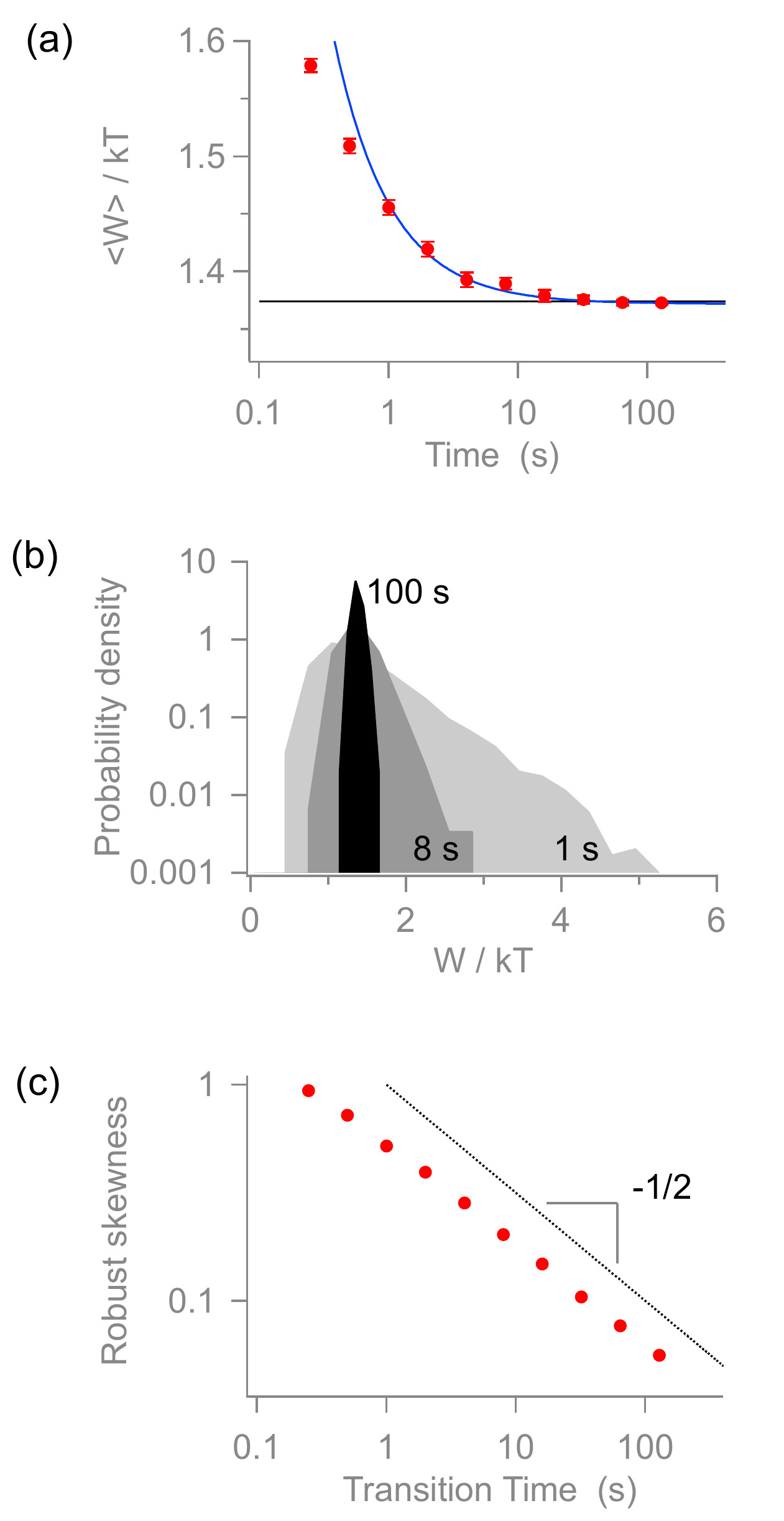}
	\caption{(Color online.)  Work required to vary a harmonic potential, with  $\alpha$ varying between $\alpha_i=0.05$ and $\alpha_f=0.4$ for $t_d=t_s=0.01 s$ and $D=2~\mu$m$^2$/s.  (a)  Average work as a function of transition time $\tau$.  Solid line is a fit to a $\tau^{-1}$ correction for $\tau > 1$ s.  (b)  Histogram of work measurements for $\tau = 1$, 8, and 100 s.  (c)  Robust skewness vs. $\tau$.}
\label{fig:kchangeWork}
\end{figure}

Figure~\ref{fig:kchangeWork} illustrates the work calculations and shows the approach to steady-state behavior for finite cycle time $\tau$.  In Fig.~\ref{fig:kchangeWork}(a), we see the approach to the value (horizontal black line) calculated from Eq.~\eqref{eq:work3}.  The solid curve is a fit, for large $\tau$ ($ > 1$), to the extra (dissipated) work $W_d$ that exceeds the asymptotic value   calculated in Eq.~\eqref{eq:work3}.  The fit is to the form $W_d \sim \tau^{-1}$, as suggested by Sekimoto \cite{sekimoto10}.

In Fig.~\ref{fig:kchangeWork}(b), we see that the distribution of work has a pronounced positive skew for finite transition times $\tau$ but approaches a Gaussian distribution as $\tau \to \infty$ (as expected in general for continuous systems \cite{speck04}).  A positive skew in the work distribution has been seen before in experimental measurements based on optical tweezer traps \cite{blickle06}, but a quantitative theory is so far not available.  It is interesting that the skew predicted here, which should be accessible experimentally, is much greater than that observed in \cite{blickle06}.

In Fig.~\ref{fig:kchangeWork}(c), we see that the skew decreases as $\tau^{-1/2}$ for large $\tau$.  To reduce the effect of rare outliers, we use a robust measure of skew, $[(W_{75}-W_{50})-(W_{50}-W_{25})]/W_{50}$, where $W_{50}$ is the median value of work and $W_{25}$ and $W_{75}$ represent the first and third quartile values \cite{robustskew}.  Skewness thus serves to measure disequilibrium.  An alternative measure---requiring more data to calculate with equivalent precision---is based on the relative entropy between forward and backward processes \cite{kawai07}.  The origin of the scaling of the skew measure deserves more investigation.


We note that the values for the asymptotic work differ from the ``naive" calculation of $\tfrac{1}{2} \ln \tfrac{\alpha_f}{\alpha_i}$ by $\mathcal{O}(\alpha)$ for $\alpha \ll 1$.  Since the work depends on $\langle x^2 \rangle$ and since this quantity also differs by $\mathcal{O}(\alpha)$ from the continuous value for small $\alpha$, the observation is expected.  The main point is that there are no secular terms whose error grows with $\tau$.  Rather, no matter how long the measurement, the errors are controlled by the discretization size $\alpha$.  

We also note that the work we calculate is greater than that for the corresponding continuous case.  In a recent paper, Sivak et al.~have pointed out that the integration scheme itself  injects extra ``shadow" work into the system~\cite{sivak_arxiv}.  In principle, integration schemes that are more sophisticated than Euler integration could be used to remove the effects of shadow work.

Finally, we have explored numerically the effects of a camera exposure $t_c$ and observation noise $\chi$ and find that they have only a small numerical effect (data not shown).  Such a result is expected from the results of our calculations of power spectra and variance in Section~\ref{sec:equiv2cont}, where we saw that $t_c$ and $\chi$ made only minor shifts in those quantities.

\section{Discussion and Conclusions}
\label{sec:discussion}

In this article, we have shown that the complications brought on by finite time scales in virtual potentials can be accounted for and do not lead to radical differences between virtual and actual potentials, as long as we are careful.  We find that  calculations of dynamics---such as variance, power spectrum, and work---lead to answers that agree with continuum calculations to $\mathcal{O}(\alpha)$, where $\alpha$ is the ratio of the feedback update time to the relaxation time of the potential.  The situation is then similar to that of ordinary experiments that also discretize continuous signals.  The main practical difference is that experiments with a physical potential can usually sample at very small values of $\alpha$ (e.g., $\alpha = 3\times 10^{-4}$ in \cite{imparato07}), whereas feedback experiments, which must act in real time on each observation, often use higher values (e.g., $\alpha = 0.1$ in \cite{cohen05b}).  As we saw in Section~\ref{sec:harmonic-perfect}, minimizing the delay helps to increase the value of the instability gain $\alpha^*$ and, by extension, to increase the range of acceptable feedback gains $\alpha$.

Although feedback traps use relatively large values of $\alpha$, they can impose more general and better-controlled motion (via virtual potentials) on a particle than can be done with physical potentials.  For example, in exploring the Landauer principle with the feedback trap, we have been able to implement the scheme proposed by Dillenschneider and Lutz \cite{dillenschneider09}, which is based on a specific parametrization of a double-well potential where the difference in well depth and barrier heights are separately controlled.  Although such manipulations can be approximately done by placing two optical traps side by side and adding a sideways flow \cite{berut12}, more general and better-controlled variations are possible in the feedback trap, since the form of the imposed potential is now arbitrary, as long as the feedback gain is small enough.

One assumption in the work presented above is that the electrokinetic force in an ABEL trap responds instantaneously to the command signal.  This is usually a good approximation: both electrophoretic and electroosmotic forces depend on a realignment of the double-layer, which is typically much less than a micron in scale.  Typical response times are then of order 10 $\mu$s \cite{cohen06}, which is much faster than the ms time scale of feedback traps.  In related work, we have used piezoelectric translation stages to \textit{track} the motion of small particles rather than trap them.  The formalism that is required is similar to that explored here, but one must then account for the finite relaxation time, of order ms, of the translation stages~\cite{schertel_thesis,schertel_unpub}.

We have also shown that calculations of work done by the virtual potential are relatively straightforward, although many theoretical questions deserve more exploration.  However, the situation is less clear regarding the calculation of heat dissipation into a thermal bath.  A naive discretization analogous to that used for work, Eq.~\eqref{eq:work0} gives the correct result when applied to harmonic potentials but has spurious secular terms that grow linearly with the protocol time for anharmonic potentials.  However, we observe similar effects when using the Euler discretization to calculate heat dissipation for ordinary Langevin equations with anharmonic potentials (but no feedback).  Indeed, how to calculate the heat properly from sampled data---with or without feedback---remains an issue of current discussion~\cite{sekimoto10,sivak_arxiv,kanazawa12}.

One final question is whether experiments can impose conditions and forces that are well-enough controlled to apply the theory described in this work. That is, we have assumed that the particle moves only because of thermal fluctuations and because of deliberately applied forces.  Knowing their effects requires a calibration of both electrical mobility (the charge on a particle) and diffusion constant (the size of a particle).  It also assumes that the electric field is known and there are no mechanical drifts.  In preliminary experimental work, we have established that it is possible to accurately describe motions at frequencies above 1 Hz, but at lower frequencies, mechanical drifts from temperature variations become important.

\vspace{2em}
\section*{Acknowledgments}
We thank Udo Seifert, Paul Tupper, Jordan Horowitz, David Sivak, and the two referees for helpful comments.  This work was supported by NSERC (Canada).

\section*{Appendix}
We prove Eq.~\eqref{eq:lemma}, which is used in the calculation of average work.
\begin{lem}
\begin{equation}
	\lim_{\tau \to \infty} t_s \left( \frac{\alpha_f-\alpha_i}{\tau} \right) 
	\sum_{n=0}^{\tau/t_s}
	\frac{1}{b+\alpha_n} = \ln \left( \frac{b+\alpha_f}{b+\alpha_i} \right) \,.
\label{eq:lemma1}
\end{equation}
\end{lem}
\textit{Proof}.  Intuitively, the expression is simply $\int_{\alpha_i}^{\alpha_f} \tfrac{d\alpha}{b+\alpha}$.  More formally, with $\Delta \alpha = \alpha_f - \alpha_i$ and $N=\tau/t_s$, we have 
\begin{align}
	&\frac{\Delta \alpha}{N} 
	\sum_{n=0}^{N} \frac{1}{b+\alpha_i+\frac{\Delta \alpha n}{N}} 
	= \sum_{n=0}^{N} \frac{1}{\frac{(b+\alpha_i)N}{\Delta \alpha}+n}
		\nonumber \\[3pt]
	=& \psi \left(1+\frac{(b+\alpha_i)}{\Delta \alpha}N+N \right) - 
	\psi \left( \frac{(b+\alpha_i)N}{\Delta \alpha} \right) \,.
\end{align}
The digamma function $\psi(x) \sim \ln x$ for $x \to \infty$.  Taking $N \to \infty$, we have
\begin{align}
	\ln \left( \frac{1+\frac{(b+\alpha_i)N}{\Delta \alpha}+N}{\frac{(b+\alpha_i)N}{\Delta \alpha}} \right) &\approx
	\ln \left(\frac{\frac{b+\alpha_i}{\Delta \alpha}+1}
	{\frac{(b+\alpha_i)}{\Delta \alpha}} \right) \nonumber \\
	&= \ln \left(1+\frac{\Delta \alpha}{b+\alpha_i} \right) \nonumber \\
	&= \ln \left(\frac{b+\alpha_f}{b+\alpha_i} \right) \,.
\end{align}



\begin{thebibliography}{20}

\bibitem{cohen05a}  A.~E.~Cohen and W.~E.~Moerner, 
App.~Phys.~Lett. \textbf{86}, 093109 (2005).

\bibitem{fields11} A.~P.~Fields and A.~E.~Cohen, Proc.~Nat.~Acad.~Sci. (PNAS) \textbf{108} 8937 (2011).

\bibitem{cohen05b} A.~E.~Cohen, 
Phys.~Rev.~Lett. \textbf{94}, 118102 (2005).

\bibitem{landauer61}  R.~Landauer, IBM J.~Res.~Develop. \textbf{5}, 183 (1961).

\bibitem{berut12} A.~B\'erut et al., Nature \textbf{483}, 187 (2012).

\bibitem{cho11} A.~Cho, Science \textbf{332}, 171 (2011).

\bibitem{kubo66} R.~Kubo, Rep.~Prog.~Phys. \textbf{29}, 255 (1966).

\bibitem{marconi08} U.~M.~B.~Marconi, A.~Puglisi, L.~Rondoni, and A.~Vulpiani, Phys.~Rep. \textbf{461} 111 (2008).

\bibitem{cohen06} A.~E.~Cohen, 
Ph. D. Thesis, Stanford Univ., 2006.

\bibitem{goulian00} M.~Goulian and S.~M.~Simon, 
Biophys.~J. \textbf{79}, 2188 
(2000).

\bibitem{savin05} T.~Savin and P.~S.~Doyle, 
Biophys.~J. \textbf{88}, 623 
(2005).

\bibitem{gardiner09} C.~Gardiner, \textit{Stochastic Methods}, 4th ed. (Springer-Verlag, Berlin, 2009).

\bibitem{astrom97} K. J. \r{A}str\"{o}m and B. Wittenmark, \textit{Computer-Controlled Systems: Theory and Design}, 3rd ed. (Prentice Hall, 1997).  Reprinted by Dover Books, 2011.

\bibitem{bechhoefer05} J.~Bechhoefer, Rev.~Mod.~Phys.~\textbf{77}, 783 (2005).

\bibitem{gosse02}  C.~Gosse and V.~Croquette, 
Biophys.~J. \textbf{82}, 3314 
(2002).

\bibitem{west03} B.~J.~West, M.~Bologna, and P.~Grigolini, \textit{Physics of Fractal Operators} (Springer, 2003).

\bibitem{sekimoto97} K.~Sekimoto, 
J.~Phys.~Soc.~Japan \textbf{66}, 1234 
(1997).

\bibitem{sekimoto10} K.~Sekimoto, \textit{Stochastic Energetics} (Springer-Verlag, Berlin, 2010).

\bibitem{blickle06} V.~Blickle, T.~Speck, L.~Helden, U.~Seifert, and C.~Bechinger, Phys.~Rev.~Lett. \textbf{96}, 070603 (2006).

\bibitem{schmiedl07} T.~Schmiedl and U.~Seifert, 
Phys.~Rev.~Lett. \textbf{98}, 108301 (2007).

\bibitem{speck04} T.~Speck and U.~Seifert, 
Phys.~Rev.~E \textbf{70}, 066112 (2004).

\bibitem{robustskew}  Robust skewness is often defined normalizing not by $W_{50}$ but instead by $W_{75}-W_{25}$.

\bibitem{kawai07} R.~Kawai, J.~M.~R.~Parrondo, and C.~Van den Broeck, 
Phys.~Rev.~Lett. \textbf{98}, 080602 (2007).


\bibitem{sivak_arxiv} D.~A.~Sivak, J.~D.~Chodera, and G.~E.~Crooks, arXiv.org 1107.2967v3.

\bibitem{imparato07} A.~Imparato, L.~Peliti, G.~Pesce, G.~Rusciano, and A.~Sasso, Phys.~Rev.~E \textbf{76} 050101(R) (2007).

\bibitem{dillenschneider09} R.~Dillenschneider and E.~Lutz, Phys.~Rev.~Lett. \textbf{102}, 210601 (2009).

\bibitem{schertel_thesis} L.~Schertel, Bachelor's Thesis, Univ. of Konstanz, Germany (2012).

\bibitem{schertel_unpub} L.~Schertel, D.~Wiedmann, and J.~Bechhoefer, unpublished.

\bibitem{kanazawa12} K.~Kanazawa, T.~Sagawa, and H.~Hayakawa, Phys.~Rev.~Lett \textbf{108}, 210601 (2012).

\end{thebibliography}
\end{document}